# Physics-Informed Neural Networks for microflows: Rarefied Gas Dynamics in Cylinder Arrays

Jean-Michel Tucny[a,b*], Marco Lauricella[c], Mihir Durve[a], Gianmarco Guglielmo[b], Andrea Montessori[b], Sauro Succi[a,d]

[a]Center for Life Nano- & Neuro-Science, Fondazione Istituto Italiano di Technologia (IIT), viale Regina Elena 295, 00161 Rome, Italy

[b]Dipartimento di Ingegneria Civile, Informatica e delle Tecnologie Aeronautiche, Università degli Studi Roma Tre, via Vito Volterra 62, Rome, 00146, Italy

[c]Istituto per le Applicazioni del Calcolo del Consiglio Nazionale delle Ricerche, via dei Taurini, 19, Roma, 00185, Italy

[d]Department of Physics, Harvard University, 17 Oxford St. Cambridge, MA 02138, United States

*Corresponding author e-mail: jeanmichel.tucny@uniroma3.it

Contributors' emails: m.lauricella@iac.cnr.it, mihir.durve@iit.it, gianmarco.guglielmo@uniroma3.it, andrea.montessori@uniroma3.it, sauro.succi@iit.it

**Abstract.** Accurate prediction of rarefied gas dynamics is crucial for optimizing flows through microelectromechanical systems, air filtration devices, and shale gas extraction. Traditional methods, such as discrete velocity and direct simulation Monte Carlo (DSMC), demand intensive memory and computation, especially for microflows in non-convex domains. Recently, physics-informed neural networks (PINNs) emerged as a meshless and adaptable alternative for solving non-linear partial differential equations. We trained a PINN using a limited number of DSMC-generated rarefied gas microflows in the transition regime ($0.1 < Kn < 3$), incorporating continuity and Cauchy momentum exchange equations in the loss function. The PINN achieved under 2% error on these residuals and effectively filtered DSMC's intrinsic statistical noise. Predictions remained strong for a tested flow field with $Kn = 0.7$, and showed limited extrapolation performance on a flow field with $Kn = 5$ with a local overshoot of about 20%, while maintaining physical consistency. Notably, each DSMC field required ~20 hours on 4 graphics processing units (GPU), while the PINN training took $< 2$ hours on one GPU, with evaluations under 2 seconds.

## 1. Introduction

Rarefied gas flows occur when the molecular mean free path $\lambda$ is comparable to the characteristic length $L_C$ of the flow, which is common in microelectromechanical systems [1, 2], air filtration devices [3] and shale gas extraction [4]. In these applications, the propagation of gas molecules and the transport of physical parameters are coupled, rendering the assumptions of continuum mechanics invalid. The degree of rarefaction of gas flows is characterized by the Knudsen number ($Kn = \lambda/L_C$). While the Boltzmann equation accurately represents the evolution of the probability distribution function for all Knudsen numbers [5-7], exact analytical solutions exist only for a few simple geometries. The transition regime ($0.1 < Kn < 10$) poses the greatest challenge, as both gas-gas molecular collisions and gas-solid collisions influence the flow [8]. Two key phenomena occur in this regime: (1) slip, or non-zero velocity at the gas-solid boundary, and (2) a breakdown in the linear relationship between stress and shear within the Knudsen layer, which extends $\sim 2\lambda$ from solid surfaces. These effects become more complicated with realistic porous media with non-convex gas domains. For instance, around a cylinder of diameter $D$, a secondary Knudsen layer (S-layer) emerges due to discontinuities in the probability distribution function at grazing angles [6, 9-11]. Thus, rarefied gas flows pose significant mathematical challenges in industrial applications.

Conventional numerical methods for discretizing the Boltzmann equation include the discrete velocity method (DVM) and the discrete ordinate method (DOM) [12-15], which use numerical stencils to represent velocity space. The more recently proposed discrete unified gas kinetic scheme (DUGKS) allows the use of adaptive, non-structured grids [16-20]. However, the required stencil size increases with the Knudsen number, making it difficult to determine the appropriate size beforehand, especially in multiscale porous media, leading to high computational costs. Alternatively, the Direct-Simulation Monte Carlo (DSMC) tracks the collision and propagation of random molecular samples [21]. It is widely used for high-speed rarefied gas flows in spacecraft re-entry [22-27]. However, in low-speed flows, a large number of molecular collisions is needed to reduce statistical noise, necessitating long averaging times. Despite these limitations, the DSMC has been successfully applied to laminar and weakly compressible flows past a cylinder [28-32].

Recently, deep learning has emerged as a promising alternative for solving non-linear partial differential equations (PDEs) in fluid mechanics [33-36]. These methods are particularly effective when boundary conditions (BCs) or other transport properties are incomplete, as neural networks can implicitly encode such information from sparse data. Governing equations formulated as PDEs boundary values can be seamlessly integrated in the loss function, allowing the network to serve as a universal approximator [37]. This approach, known as physics-informed neural networks (PINNs), is meshless, highly adaptable and leverages automatic differential techniques for graphics processing unit (GPU) acceleration.

PINNs have been studied for flows around cylinders [38, 39], but these studies focused on continuum regimes with, no-slip boundary conditions and uniform viscosity. PINNs have also been used for rarefied gas flows with different loss functions, including the DVM for thermal creep flow [40], Poiseuille flow [41], a triangular duct and a cavity [42, 43]. Moment methods have been applied to shock problems with periodic boundary conditions, where Galilean invariance was enforced by generalized moments [44] or convolutional neural networks [45]. Loss functions based on high-order spatial derivatives have also been explored [46], and recent efforts attempted to recover constitutive relationships in a slit [47] and in a 1-D stationary shock [48, 49]. However, all these studies featured convex flow domains, and did not address the discontinuities in probability distribution functions and their moments in non-convex geometries. Furthermore, in practice, the training process of a PINN is sensitive not only to design choices such as activation functions, learning rates and weights on components of the loss function [50, 51], data noise and gradient behavior [52, 53]. To our knowledge, no studies have addressed PINNs for rarefied gas flows with non-convex geometries.

In this paper, we use the Cauchy momentum equation as the physics-informed component of the loss function. We demonstrate this approach by training a PINN on rarefied gas flows perpendicular to an array of cylinders, using a DSMC-generated data set over a wide range of Knudsen numbers. The remainder of the paper is divided as follows. The problem is formulated in Section 2, taking into consideration DSMC computational constraints, detailed in Section 3. Our PINN design will be outlined in Section 4, with results discussed in Section 5. Concluding remarks are presented in Section 6.

## 2. Problem formulation

It is aimed to compute a steady-state, laminar and weakly compressible rarefied gas flow flowing through a two-dimensional array of cylinders, approximated as an infinitely repeating pattern with a superficial velocity $U$, as illustrated in Fig. 1. The thermodynamic properties of the gas, including temperature $T$ density $\rho$ and pressure $P$, are assumed to show negligible variation throughout the array. Under these conditions, we only need to focus on the flow through one unit cell.

The Knudsen number $Kn$ is defined as the ratio of the mean free path of gas molecules $\lambda$, to the diameter of the cylinders $D$, which we chose as the characteristic length as follows:

$$Kn = \frac{2\lambda}{D} \qquad (1)$$

where according to the perfect gas law:

$$P = \rho RT \qquad (2)$$

the mean free path of the gas molecules can be approximated by the following relationship [6]:

$$\lambda = \frac{\mu}{P}\sqrt{\frac{\pi RT}{2}} \qquad (3)$$

and where $\mu$ is the viscosity of the gas at temperature $T$ and $R$ is the specific perfect gas constant associated to the gas species. As most micro/nano flows are in the laminar regime, it is aimed to keep a Reynolds number $Re$ below 1:

$$Re = \frac{\rho UD}{\mu} < 1 \qquad (4)$$

For an ideal gas, the Mach number of the incoming gas is defined as the ratio of the superficial velocity and the speed of sound of the gas as follows:

$$Ma = \frac{U}{\sqrt{\gamma RT}} \qquad (5)$$

where $\gamma$ is the ratio of specific heat capacities. For diatomic nitrogen, $\gamma = 1.4$. A relationship between the Knudsen, Reynolds and Mach numbers can be readily obtained:

$$Ma = \sqrt{\frac{2}{\gamma\pi}} Re \cdot Kn \tag{6}$$

As a great variety of micro/nano flows are only weakly compressible, it is also aimed to keep $Ma < 0.1$, thereby limiting compressibility effects.

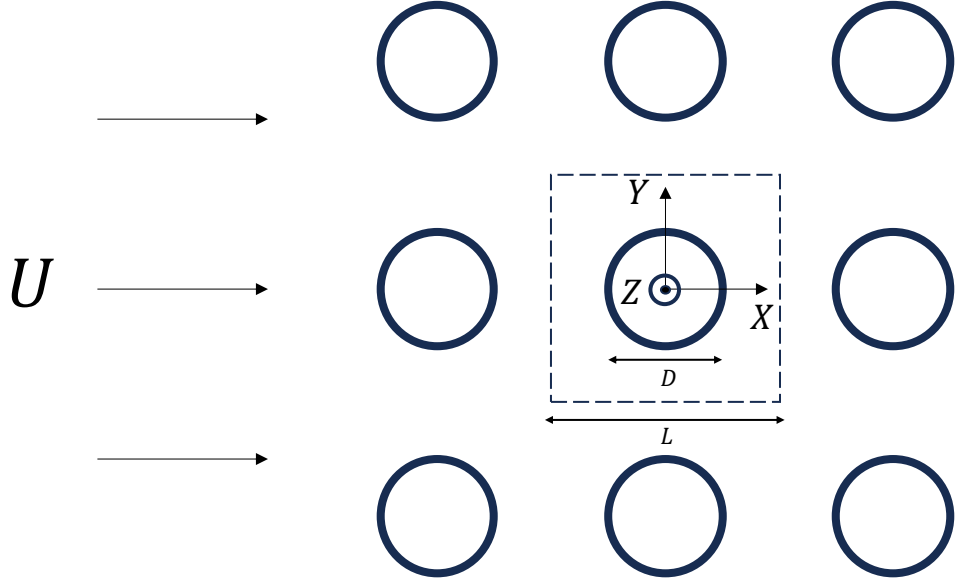

Fig. 1 – Schematic representation of the domain of the rarefied gas flow through a periodic cylinder array.

Integrating the Boltzmann equation in velocity space yields the continuity equation. Under the steady-state assumption and neglecting compressibility effects, the equation simplifies to [5, 6]:

$$0 = \nabla \cdot \mathbf{v} \tag{7}$$

Additionally, assuming laminar flow, Cauchy's momentum equation for rarefied gas can be formulated as [5, 6]:

$$0 = -\nabla p - \nabla \cdot \boldsymbol{\tau} - \rho \nabla \cdot (\mathbf{v} \cdot (\mathbf{v})) + \rho \boldsymbol{a} \tag{8}$$

where $\nabla p$ is the pressure gradient, $\rho \boldsymbol{a}$ is the external driving force (density multiplied by acceleration $\boldsymbol{a}$), and $\boldsymbol{\tau}$ denotes the deviatoric stress tensor. In two dimensions, $\boldsymbol{\tau}$ can be expressed as:

$$\boldsymbol{\tau} = \begin{bmatrix} \tau_{xx} & \tau_{xy} \\ \tau_{yx} & \tau_{yy} \end{bmatrix} \tag{9}$$

Due to the conservation of linear and angular momentum, the deviatoric stress tensor is symmetric, i.e. $\tau_{ij} = \tau_{ji}$, reducing the independent components to three in two-dimensional flow.

The computational domain size $L$ is linked to the solid fraction $\beta$ of the cylinder array:

$$L = D\sqrt{\frac{\pi}{4\beta}} \tag{10}$$

In all simulations, $L/D = 4$. From Eq. (10), it follows that $\beta \approx 0.05$.

With an infinitely repeating domain, periodic boundary conditions are applied to all flow variables at the domain boundaries. On the cylinder surfaces, we assume fully diffuse gas-wall interactions (i.e Maxwellian). While macroscopic descriptions of this phenomenon are feasible for planar flows [54], modeling this kinetic phenomenon around complicated geometries remains challenging [55]. Alternatively, a non-penetration boundary condition can be written as:

$$0 = \mathbf{v} \cdot \mathbf{n} \qquad (11)$$

Although these boundary conditions are specified for a two-dimensional cylinder array, they can be easily extended to three-dimensional domains.

## 3. Data generation with the Direct Simulation Monte-Carlo

The Direct Simulation Monte-Carlo (DSMC) method is a robust technique for simulating rarefied gas flows in the transition regime, making it suitable for generating data for training and testing the PINN in this study. DSMC models a gas as a collection of discrete sample molecules that propagate and undergo collisions, thus serving as a Lagrangian representation of the Boltzmann equation [21]. The versatility of DSMC lies in its applicability to various flow regimes, ranging from continuum to rarefied gas flows, as well as laminar, turbulent, thermal and supersonic flows [56]. However, in the continuum limit (i.e. when $Kn \ll 1$), DSMC requires resolving numerous collisions for accurate flow depiction, leading to increased computational demands. Additionally, for small flow velocities, statistical noise from molecular sampling dan introduce significant uncertainties in the flow variables. These constraints are a rationale for the training of PINNs on a limited number of DSMC simulations.

For this study, SPARTA, an open-source DSMC code developed by Sandia National Laboratories, was utilized to simulate gas flows around cylinders in laminar rarefied regimes [57]. It was previously used to simulate rarefied gas flows over isolated cylinders in turbulent slip and early transitional flows [58, 59]. SPARTA leverages the Kokkos library, a portable C++ library that supports both CPU and GPU computing [60].

In all DSMC simulations, nitrogen gas (N$_2$) was considered and the variable hard sphere (VHS) collision model employed for molecular interactions. A data set was constructed over a range of Knudsen numbers from 0.1 to 5. Numerical parameters ensure the accuracy of the simulations, with a regular grid of $n_x \times n_y = 256 \times 256$ cells in the x- and y-directions. This setup guaranteed that the ratio of the mean free path to the regular grid size $\Delta x$ remained below 3 across the entire dataset, ensuring sufficient spatial resolution. The timestep was defined as:

$$\Delta t = \frac{L}{5(\bar{c} + U)n_x} \qquad (12)$$

where $\bar{c}$ is the average molecular speed. To represent, the gas flow, 6.6 million particles were simulated, which at least 100 particles in every cell. All these criteria meet or exceed established standards for DSMC simulations on rarefied gas flows [57, 58].

To maintain laminar and weakly compressible flow conditions, a uniform acceleration was applied as:

$$a_x = 0.00766 \cdot \frac{2RT}{L} \cdot Kn \qquad (13)$$

where 0.00766 is dimensionless constant and $a_x = \|\mathbf{a}\|$ in the given geometry. This scaling ensures that the force term $\rho\mathbf{a}$ remains constant in the momentum equation (Eq. (8)), but the validity of these assumptions in the DSMC computations can only be ascertained *a posteriori*. Under this acceleration term, the maximum Reynolds number was approximately $Re \approx 0.13$ at $Kn = 0.1$, and the maximum Mach number was about $Ma \approx 0.072$ at $Kn = 5$, verifying that the flow remained within laminar and weakly compressible conditions across the tested Knudsen number range. The value of the dimensionless constant (0.00766) was computed *a posteriori* to ensure reproducibility for various species, temperatures and dimensional lengths.

Simulations were run for 500000 timesteps to reach a steady-state flow. The steady-state criterion was assessed by averaging velocity across the domain over 1000 timesteps and ensuring a relative error below 0.002 in the average velocity. Following this, simulations continued an additional 6.5 million timesteps to obtain time-averaged local flow variables. The deviatoric stress tensor was calculated from the pressure and second-order moment of the molecular distribution $\mathbf{\Pi}$ using the definition:

$$\boldsymbol{\tau} = \mathbf{\Pi} - p\mathbf{I} \qquad (14)$$

where $\mathbf{I}$ is the identity matrix. Each simulation was executed on a single Intel Xeon core with 4 NVIDIA A100 SXM6 GPU HBM2 units on the Leonardo cluster at CINECA, with a memory requirement of 900 MB per GPU. Each run took roughly 20 hours to complete.

## 4. Physics-informed neural networks for rarefied gas flows

The primary objective of this study is to develop a single physics-informed neural network capable of accurately approximating local macroscopic flow variables for rarefied gas microflows in a cylinder array across a wide range of Knudsen numbers. Specifically, our goal is to approximate a function with three inputs ($x$, $y$, $Kn$) and six outputs (velocity components $v_x$ and $v_y$, pressure $p$ and stress tensor components $\tau_{xx}$, $\tau_{yy}$ and $\tau_{xy}$). In our problem formulation, only three partial differential equations (PDEs) from the governing equations are directly available. The PINN leverages these physics constraints to infer the relationships between all six outputs while maintaining consistency with the underlying equations of rarefied gas dynamics.

The smooth nature of the neural network offers the potential to mitigate statistical fluctuations inherent to the DSMC representation of gas flows, particularly in microflow regimes. These fluctuations, arising from the discrete nature of molecular-level simulations, often challenge traditional interpolation techniques. By incorporating physical constraints, the PINN framework may serve as a robust alternative for filtering such noise and enabling more accurate predictions.

To achieve this, we first review the fundamentals of PINNs, focusing on how governing equations are embedded into the loss function. We then detail the pre-processing steps applied to DSMC simulation results, including data normalization. Finally, we provide specifics on the training and testing procedures of the neural network in the last subsection.

### a. Physics-informed neural networks

Physics-informed neural networks serve as universal function approximators that integrate governing physical laws, formulated as partial differential equations (PDEs), directly into the training process as loss functions to guide the network's predictions. The hidden layers of the neural networks are expressed as a composition of simple functions as follows:

$$\begin{aligned}\mathbf{z}^{(0)} &= \mathbf{x} \\ \mathbf{z}^{(k)} &= \phi\bigl(\mathbf{W}^{(k)}\mathbf{z}^{(k-1)} + \mathbf{b}^{(k)}\bigr)\end{aligned} \quad (15)$$

where $\phi(t)$ is a non-linear activation function, and $\mathbf{W}^{(k)}$ and $\mathbf{b}^{(k)}$ are the weight matrix and the bias vector of the $k$th layer, which are the trainable model parameters. Recently, it has suggested that adaptive activation functions, where the learning rate is adjusted separately for each layer, can improve performance compared to uniform activation functions across the entire network [61, 62].

Using an optimizer such as Adam [63], the weights and biases in Eq. (15) are adjusted to minimize the mean square error between the main flow variables (velocity, pressure and deviatoric stress tensor components). The physics-informed part of the neural network consists in adding the residual of the continuum and PDEs.

Periodic boundary conditions, a key aspect of our model, are imposed by introducing a truncated Fourier series layer prior to the deep neural network [64, 65]. This method, rather than adding multiple residual terms (12 in total, 6 variables in $x$ and $y$) with conflicting penalties, was chosen in our PINN, as shown in the schematic representation of Fig. 2.

To enhance model robustness and prevent overfitting, we included regularization terms in the loss function. A summary of the neural network structure and parameters is provided in Table 1.

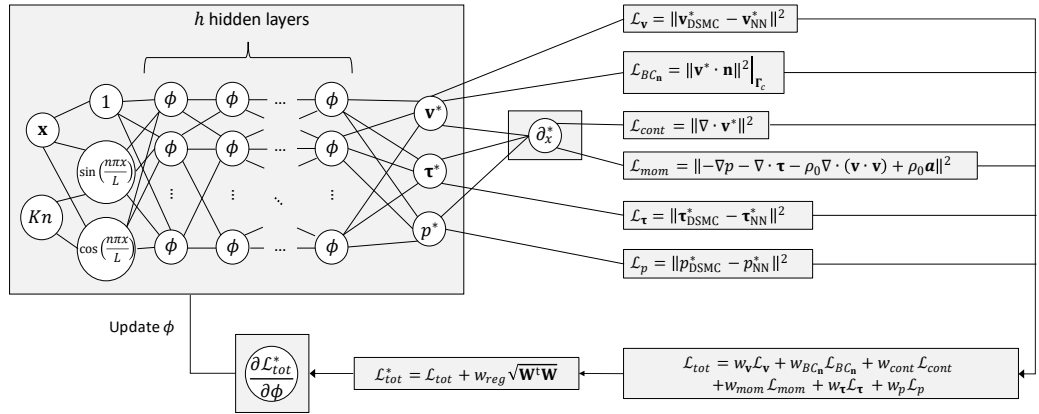

Fig. 2 – Schematic representation of PINNs for the approximation a rarefied gas flow in a cylinder array, with a layer of truncated Fourier series to impose hard periodic boundary conditions.

Table 1 – Setup of the PINN

| Hyperparameter | Value |
|---|---|
| Activation function | $\phi(t) = \tanh(t)$ |
| Weight initialization | Glorot normal |
| Optimizer | Adam |
| Weight decay | $w_{reg} = 0.001$ |
| Loss function | Mean square error |
| Weights of PDEs | $w_{cont} = w_{mom} = 0.1$ |
| Weights on data | $w_{\mathbf{v}} = w_{\boldsymbol{\tau}} = w_p = 1$ |
| Number of hidden layers | 10 |
| Number of neurons per layer | 100 |
| Number of Fourier modes | $n = 3$ |

**b. Data pre-processing through variable normalization**

Before training the PINN, input data was pre-processed to ensure consistent scaling across all variables, which is critical for effective convergence. Although conventional dimensionless variables, such as division by reference velocities, pressures and deviatoric stress tensor components seem like a natural choice for analysis, preliminary attempts have shown this choice of normalization leads to inhomogeneous data distribution. For instance, the Knudsen number varies by more than an order of magnitude across the data set, and the velocity demonstrates significant variability because of the slip phenomenon. Furthermore, in two- or three-dimensional domains, each component of the deviatoric stress tensor has inflection points that are located at different positions, with angular variations that differ from the velocity field. These problems impede the training of the PINN.

To address these challenges, we applied Z-score normalization to output variables $Y$, as typically recommended to reduce variability [66]:

$$Y^* = \frac{Y - \mu_Y}{\sigma_Y} \qquad (16)$$

where $Y^*$ is the normalized output variable, $\mu_Y$ is the mean of $Y$ on the data set and $\sigma_Y$ is the standard deviation of $Y$. However, because the Knudsen number is manipulated through the density in the data set, much of the variance in the pressure variance along the Knudsen number rather than spatial position. Therefore, pressure normalization involved subtracting the spatially averaged pressure at each Knudsen number and dividing by the standard deviation across the dataset:

$$p^*(Kn) = \frac{p - \iint_{\mathbf{x}} p(\mathbf{x}, Kn) d\mathbf{x} / \iint_{\mathbf{x}} d\mathbf{x}}{\sigma_p} \qquad (17)$$

Preliminary studies have shown that the min-max scaling to $[-1, 1]$ was functioning adequately for the position variables:

$$X^* = 2 \cdot \frac{X - \min(X)}{\max(X) - \min(X)} - 1 \qquad (18)$$

where $X^*$ is the normalized input variable. However, due to the Knudsen number's broad range of values, better results were achieved by applying a natural logarithm is used ($\ln(Kn)$) instead.

This normalization alters variable scaling (Eqs. (16)-(18)), impacting the computation of the residuals in the continuum and momentum balance equations. Additionally, it is desirable to ensure that loss terms maintain comparable magnitudes. Recalling $\rho a$ and $L$ are uniform across the data set, the loss functions were adjusted as follows:

$$\mathcal{L}_{\text{cont}} = \left\| \frac{L\left(\sigma_{v_x}\frac{\partial v_x}{\partial x} + \sigma_{v_y}\frac{\partial v_y}{\partial y}\right)}{2\sigma_{v_x}} \right\|^2 \tag{19}$$

$$\mathcal{L}_{mom_x} = \left\| \frac{-L\left(\sigma_{\tau_{xx}}\frac{\partial \tau_{xx}}{\partial x} + \sigma_{\tau_{xy}}\frac{\partial \tau_{xy}}{\partial y} + \sigma_p\frac{\partial p}{\partial x} + \left((v_x\sigma_{v_x} + \mu_{v_x})\frac{\partial v_x}{\partial x} + (v_y\sigma_{v_y} + \mu_{v_y})\frac{\partial v_x}{\partial y}\right)\sigma_{v_x}\rho_0\right)}{2\rho_0 a} + 1 \right\|^2 \tag{20}$$

$$\mathcal{L}_{mom_y} = \left\| \frac{-L\left(\sigma_{\tau_{xy}}\frac{\partial \tau_{xy}}{\partial x} + \sigma_{\tau_{yy}}\frac{\partial \tau_{yy}}{\partial y} + \sigma_p\frac{\partial p}{\partial y}\left((v_x\sigma_{v_x} + \mu_{v_x})\frac{\partial v_y}{\partial x} + (v_y\sigma_{v_y} + \mu_{v_y})\frac{\partial v_y}{\partial y}\right)\sigma_{v_y}\rho_0\right)}{2\rho_0 a} \right\|^2 \tag{21}$$

where $\rho_0(Kn)$ is the average mass density for a given Knudsen number. The multiplication by $L/2$ argument removes the gradient operator's dimensionality. Finally, since the pressure scaling induced by Eq. (17) is only a function of the Knudsen number, $\rho_0$ values can be retrieved using the perfect gas law (Eq. (2)) which is applicable here.

### c. Training and testing

The data set generated by SPARTA was divided into separate training and testing data sets, with data unused in training allocated for testing purposes. Specifically, SPARTA results with $Kn = 0.7$ and $Kn = 5$ were withheld from the training data set. The former Knudsen number was excluded to evaluate the PINN's performance where the Knudsen layer's inner and outer regions are both significant, while the latter Knudsen number was selected to assess the PINN's ability to extrapolate beyond the training Knudsen range. Afterwards, 20000 points were randomly sampled from the SPARTA generated flow fields for training on 9 Knudsen numbers in the range $0.1 < Kn < 3$, representing over 2000 points per Knudsen number.

To account for physical phenomena near the cylinder, 5000 PDE residual evaluation points were added within a radius of $0.2D$ near the cylinder surface. Additionally, 10000 additional PDE residual evaluation points were added across the entire spatial domain. Fig. 3 shows an example of the distribution of these residual evaluation points.

The neural network underwent training for 10000 iterations with a cosine decay schedule $\eta_t$ defined as follows $t$:

$$\eta_t = \eta_0 \cdot \frac{1}{2}\left(\alpha + (1-\alpha)\cos\left(\frac{\pi t}{T}\right)\right) \tag{22}$$

with $\eta_0 = 0.0001$, $\alpha = 0.1$ and $T = 10000$. The model was implemented and trained using the scientific machine learning platform DeepXDE [67], alongside TensorFlow [68] and CUDA [69].

Overall, a maximum of 3% of the data generated by SPARTA was used to train a single PINN across the wide range of Knudsen numbers, with the remaining 97% reserved for testing. Neural network training required approximately 2 hours, while evaluations took less than 2 seconds using one CPU Intel Xeon core with 1 NVIDIA A100 SXM6 GPU HBM2 on the Leonardo cluster provided by CINECA. All computations were carried on using 64-bit floating-point representations.

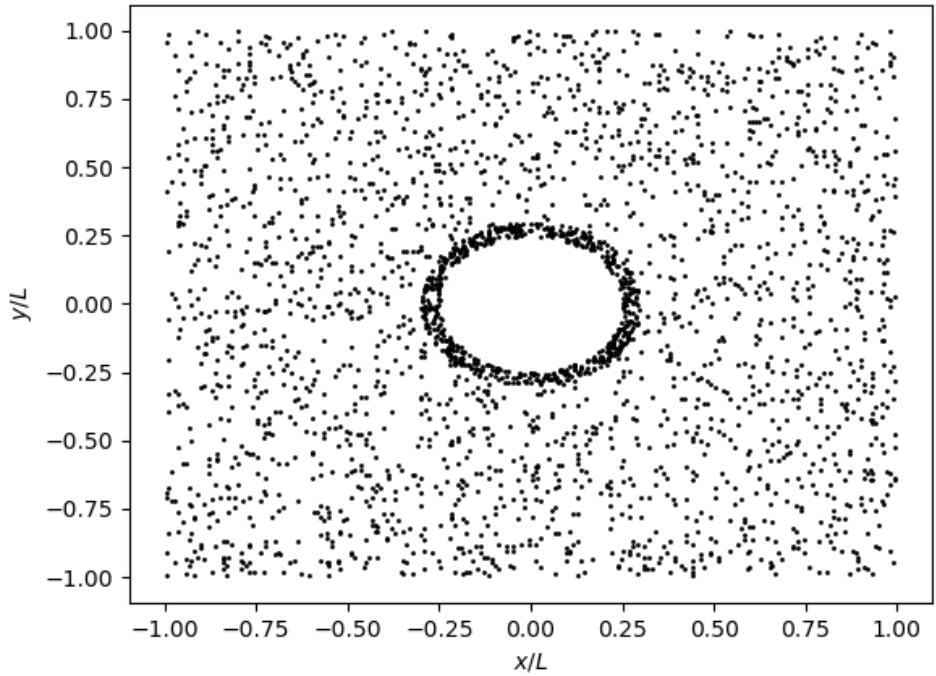

Fig. 3 – Spatial distribution of the residual evaluation points located for $0.1 < Kn < 0.15$.

## 5. Results

In this section, we demonstrate the performance of our PINN by examining selected flow field variables and assessing the source of errors and residuals. Given that the PINN was trained on DSMC data, which inherently includes statistical fluctuations, we begin by displaying the time-averaged data to illustrate the impact of noise on the gradients of flow variables. We then present the outputs of the PINN, comparing them with DSMC data both on Knudsen numbers which were used for the training and those that were reserved for both the Knudsen numbers included in the training set and those reserved for testing. Finally, we use the trained PINN to explore how the Knudsen number influences the stress-shear relationship across the rarefied gas domain.

### a. DSMC dataset

The DSMC results produced by SPARTA at the boundaries of the training dataset range ($Kn = 0.1$ and $Kn = 3$). Notably, the narrow ranges on the figures shown that the assumptions of weak compressibility and isothermality hold true. Comparing Figs. 4a and 4b, significant statistical fluctuations are observed at $Kn = 0.1$, particularly on second-order moments (notably the pressure and deviatoric stress tensors), but diminish with the Knudsen number. This trend results from a constant volumetric force term, which, while producing similar values of pressure drops and deviatoric stress tensors variations, leads to higher velocities due to the slip phenomenon at the cylinder's surface. Since statistical fluctuations are roughly proportional to mean molecular speed under these assumptions, the increased velocity reduces the relative impact of these fluctuations.

Figs. 5 and 6 display the derivatives of flow variables relevant to the computation of the continuum and momentum balances, respectively. These derivatives were obtained using a second-order centered finite difference scheme. Due to the derivative operation's tendency to amplify statistical fluctuations, noticeable fluctuations appear in the continuum balance, particularly at lower Knudsen numbers (Fig. 5). For the momentum balances (Fig. 6), the fluctuations are even more prominent due to the higher initial fluctuations in second-order moments compared to velocities before differentiation. Table 2 compiles spatial means and standard deviations of residuals from Eqs. (19)-(21), verifying that the residuals are centered around zero and are consistent with the ranges and patterns observed in Figs. 5 and 6.

Table 2 – Statistics on residuals

| $Kn$ | $\mu(\mathcal{L}_{cont})$ | $\sigma(\mathcal{L}_{cont})$ | $\mu(\mathcal{L}_{mom_x})$ | $\sigma(\mathcal{L}_{mom_x})$ | $\mu(\mathcal{L}_{mom_y})$ | $\sigma(\mathcal{L}_{mom_y})$ |
|---|---|---|---|---|---|---|
| 0.1 | 1.85E-4 | 1.67 | -0.0170 | 19.1 | 1.16E-3 | 19.1 |
| 3 | -6.86E-4 | 1.67 | -8.4E-4 | 0.611 | -5.16E-5 | 0.614 |

a) 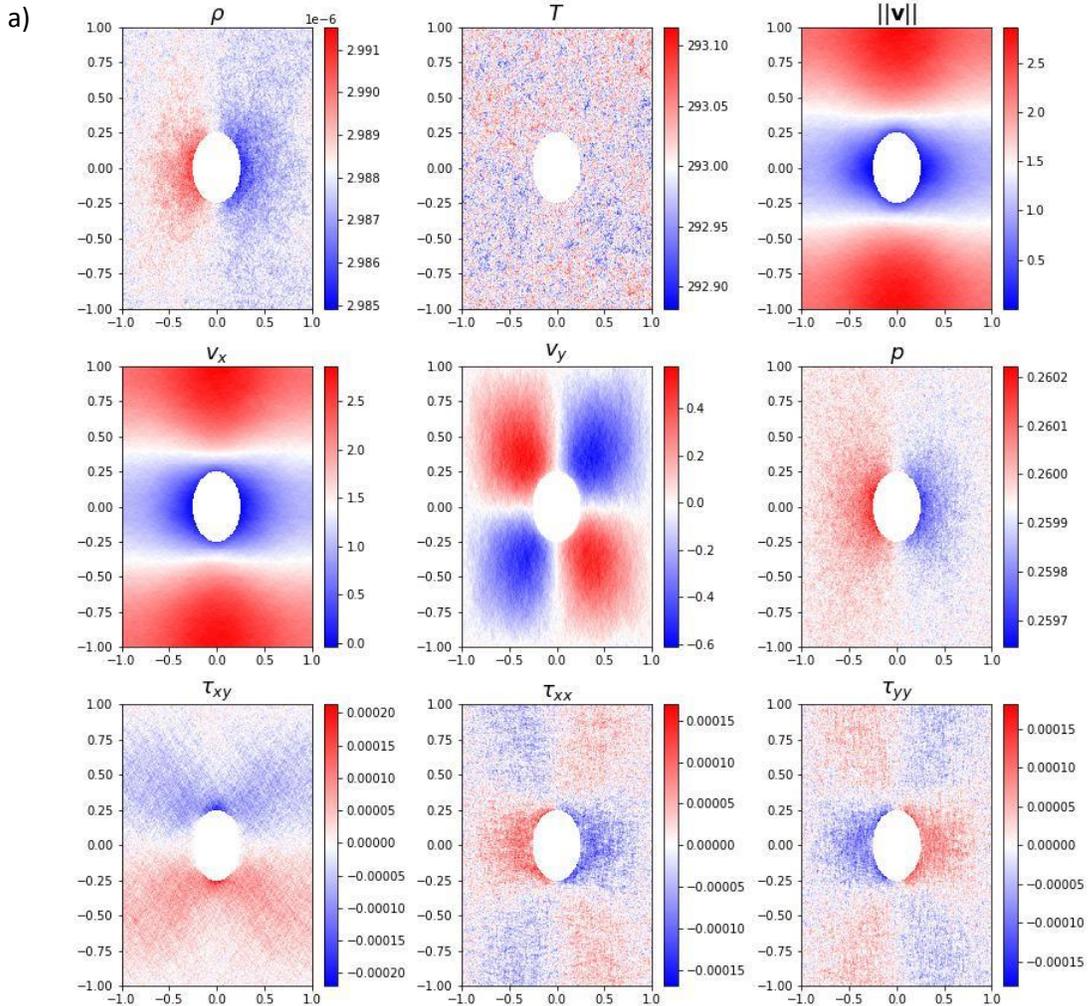

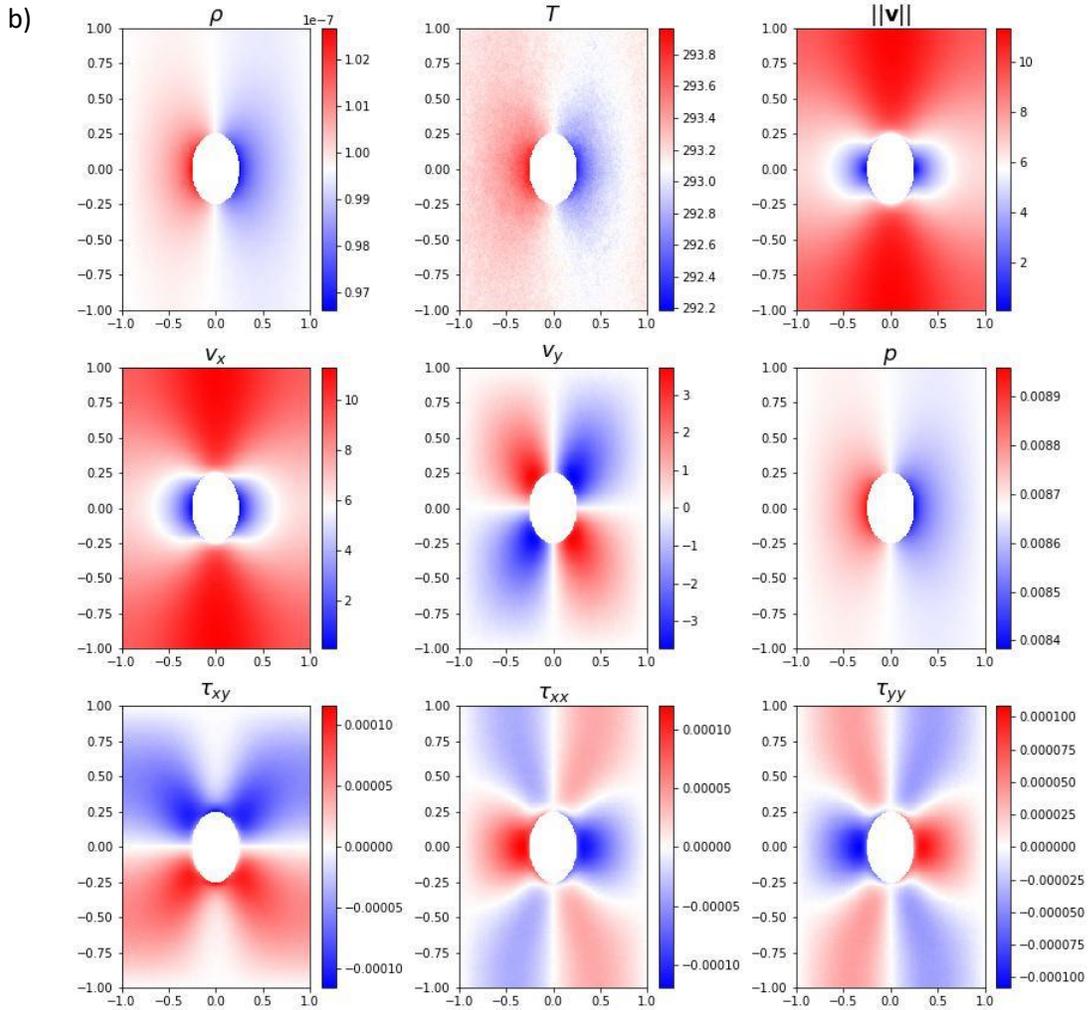

Fig. 4 – Mean fields in SI units for the flow perpendicular to a cylinder array produced by SPARTA for a) $Kn = 0.1$, b) $Kn = 3$. The flow bypasses the cylinder, which creates the y-component of the velocity field, resulting in the multiple inflection points of the deviatoric stress components. As the Knudsen number increases, slip becomes the dominant factor of the flow.

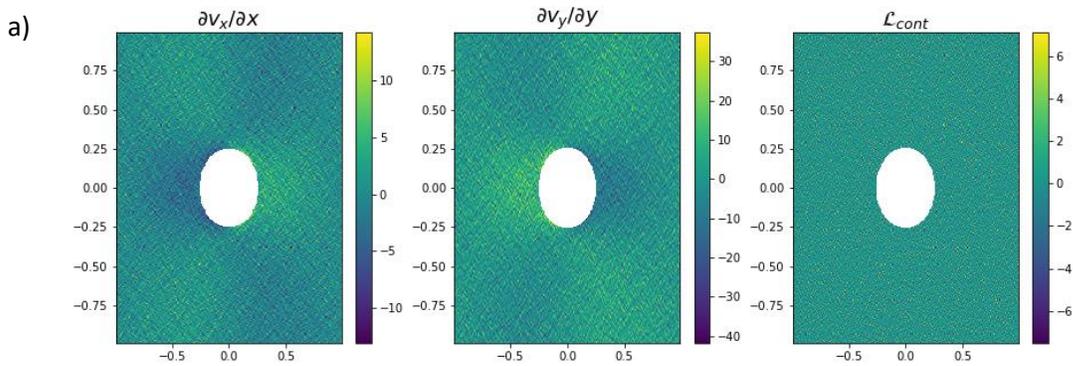

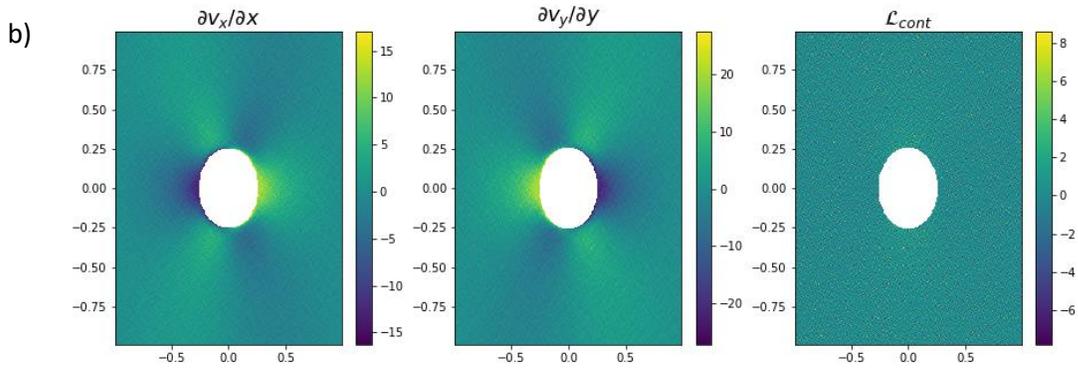

Fig. 5 – Dimensionless velocity gradients and residuals of the continuity equation of the DSMC fields for a) $Kn = 0.1$, b) $Kn = 3$. Distinctive patterns are observed for the gradients: the residual is simply zero-centered noise.

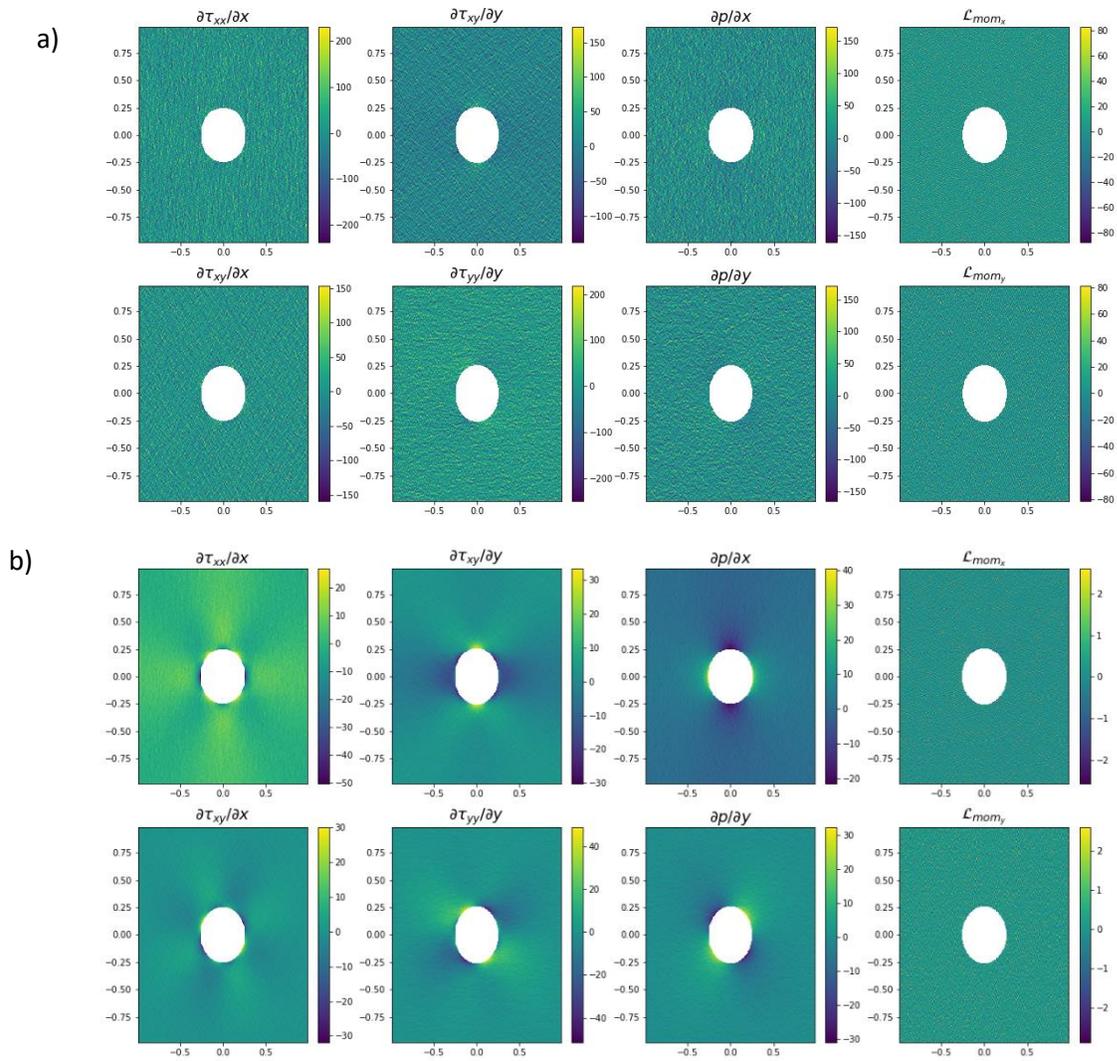

Fig. 6 – Dimensionless second-order moment gradients and residuals of the Cauchy momentum equations of the DSMC fields for a) $Kn = 0.1$, b) $Kn = 3$. The gradient patterns are dominated by noise, especially for low Knudsen numbers.

### b. PINN flow variables

The model predictions for the normalized flow fields, as defined in Eqs. (16)-(17), are displayed in Fig. 7 at both ends of the Knudsen number on the training data set. The predicted flow fields are noticeably smooth, even with the statistical fluctuations in the original data. Fig. 8 shows the local distance between the PINN predictions and the DSMC data, calculated as follows:

$$E(\mathbf{x}, Kn) = \frac{|Y^* - \hat{Y}|}{\max_{\mathbf{x}}(Y^*)} \tag{23}$$

where $\hat{Y}$ is the PINN-predicted variable. Patterns in the local distance for pressure and deviatoric stress tensor components align closely with the DSMC data shown in Fig. 4, while the error in the velocity field remains low. The normalization ensures that the variable ranges stay consistent throughout the dataset, which is essential for the convergence of the training process. Fig. 9 shows the individual loss functions throughout training, revealing three distinct phases: (1) an initial phase where the PINN primarily fits the data, with slower convergence on the PDE residual; (2) a middle phase where, after establishing an initial approximation, the PINN reduces the PDE residual, which in turn reduces the data loss; and (3) a final phase where the data residual plateaus, while the PINN continues to decrease the PDE residual slowly. The plateau is primarily due to the loss on $\tau_{xx}$ and $\tau_{yy}$, which are most affected by statistical fluctuations. Despite this, the training process demonstrates robust convergence: from the training onset, the continuity and Cauchy momentum residuals very quickly become lower than the ones computed from the second-order difference scheme on the DSMC fields.

We assesed the interpolation and extrapolation capabilities of the PINN by evaluating the flow variable outputs at $Kn = 0.7$ and $Kn = 5$, shown in Fig. 10, with the local distance illustrated in Fig. 11. For $Kn = 0.7$, the PINN performs well, and performs well and occasionally surpasses performance on training data points. However, at $Kn = 5$, while the qualitative pattern of the flow is maintained, the amplitude of the flow variables shows a maximum local overshoot of about 20%. This outcome is aligned with the general limitation of machine learning models in reliable extrapolation. In our approach, the Knudsen number acts as a "silent" variable in the physics-informed portion of the network. Thus, while the physics-informed component enforces a physically coherent flow representation, the Knudsen number's impact is not as well captured outside the training range.

The global spatial mean distance $\bar{E}(Kn)$ was computed across the full Knudsen number range as follows:

$$\bar{E}(Kn) = \iint_{\mathbf{x}} E(\mathbf{x}, Kn) d\mathbf{x} \Big/ \iint_{\mathbf{x}} d\mathbf{x} \tag{24}$$

and is shown in Fig. 12. The global spatial mean distance decreases monotonically with the Knudsen number, except in the extrapolated field for $Kn = 5$. This behavior is primarily related to the statistical noise of the DSMC data rathen than the PINN's accuracy, with the exception of the extrapolated case at $Kn = 5$.

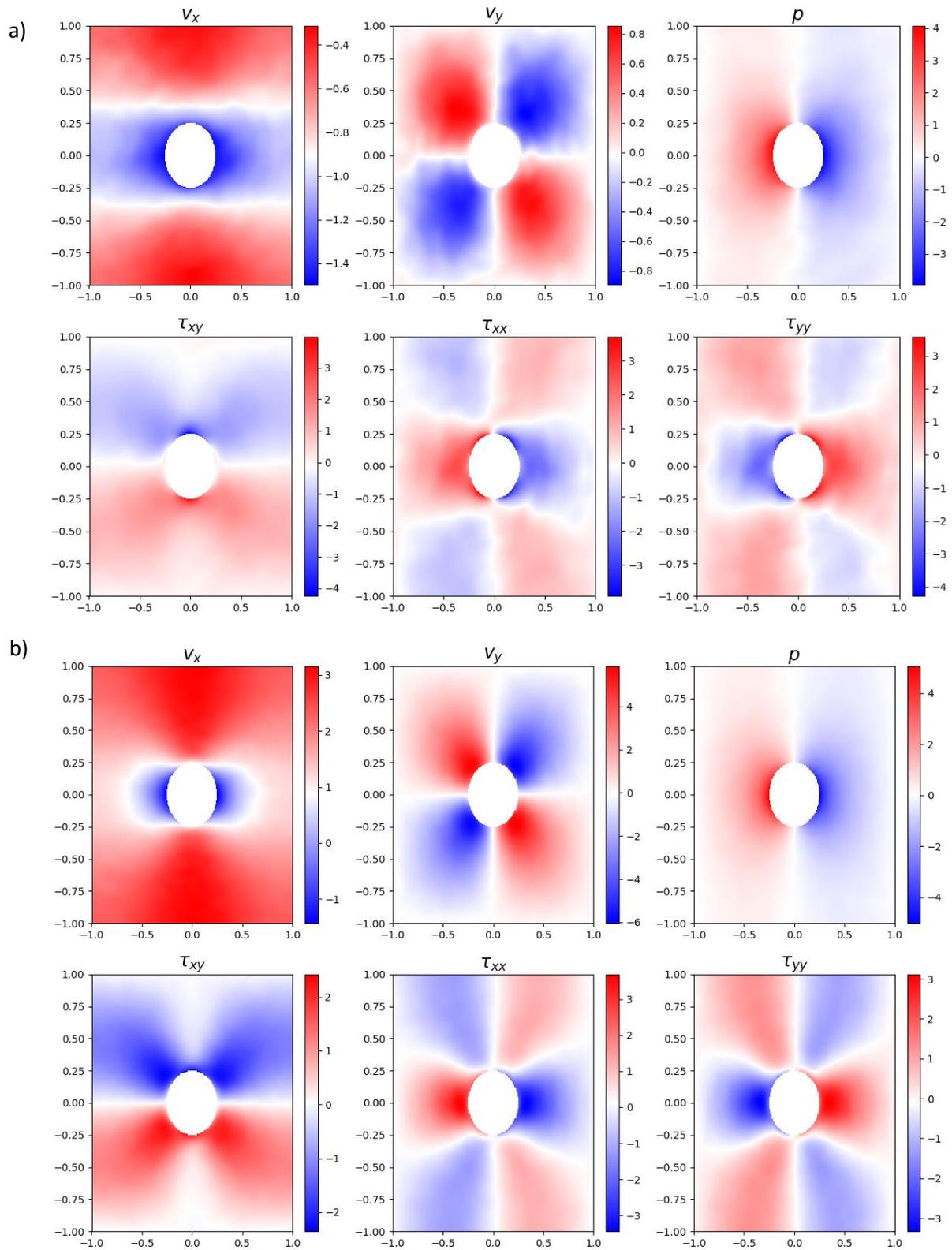

Fig. 7 – PINN predictions of normalized flow variables for a) $Kn = 0.1$, b) $Kn = 3$. Normalization was carried out on the whole dataset, which explains the negative values for $v_x$.

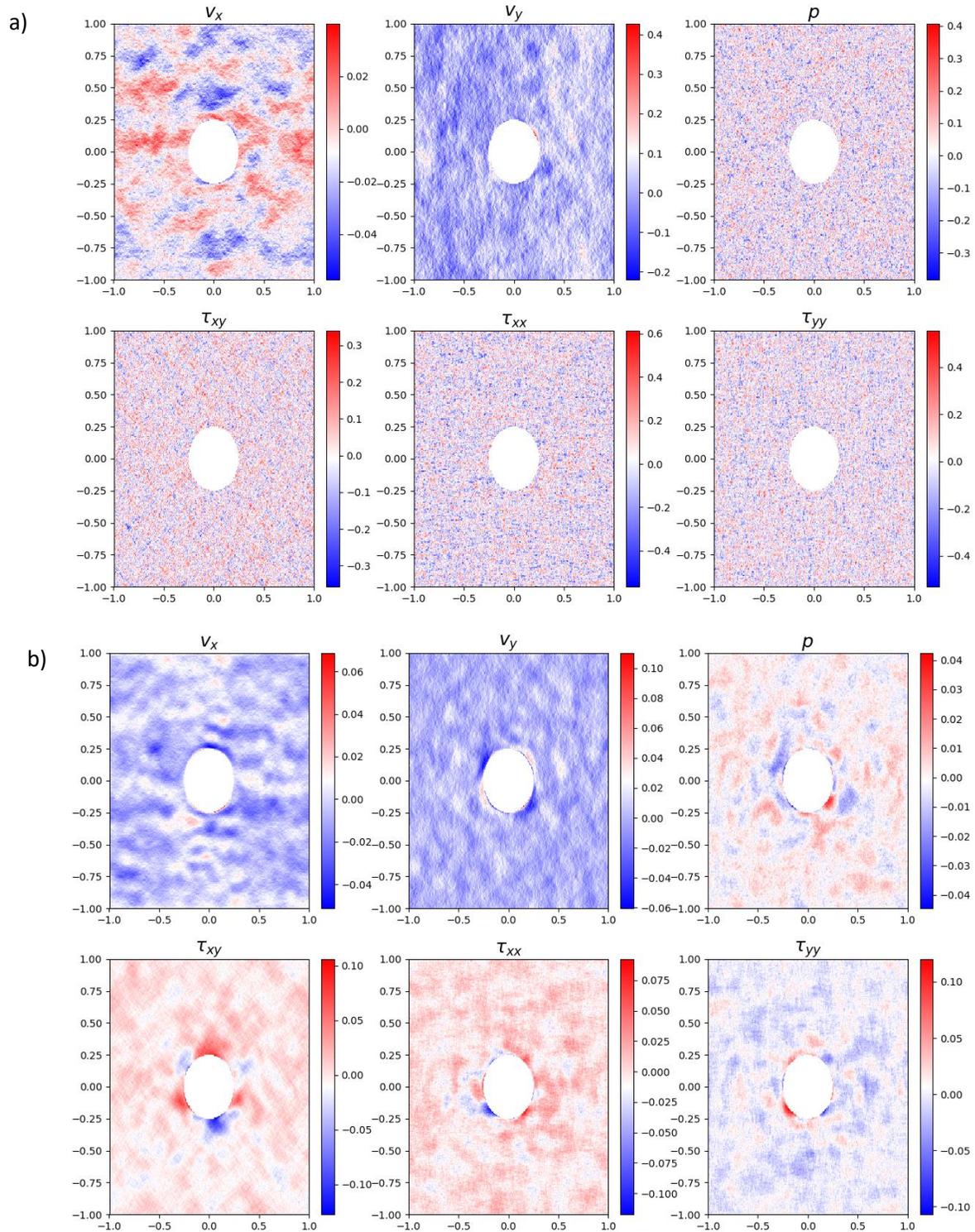

Fig. 8 – Local distance $E$ between dimensionless DSMC fields and the predictions of the PINN for a) $Kn = 0.1$, b) $Kn = 3$. At low Knudsen numbers, the local distance in pressure and deviatoric stress tensor is primarily influenced by statistical fluctuations. For high Knudsen numbers, these fluctuations diminish, with discrepancies instead arising by the complexities of inflection points at the cylinder surface.

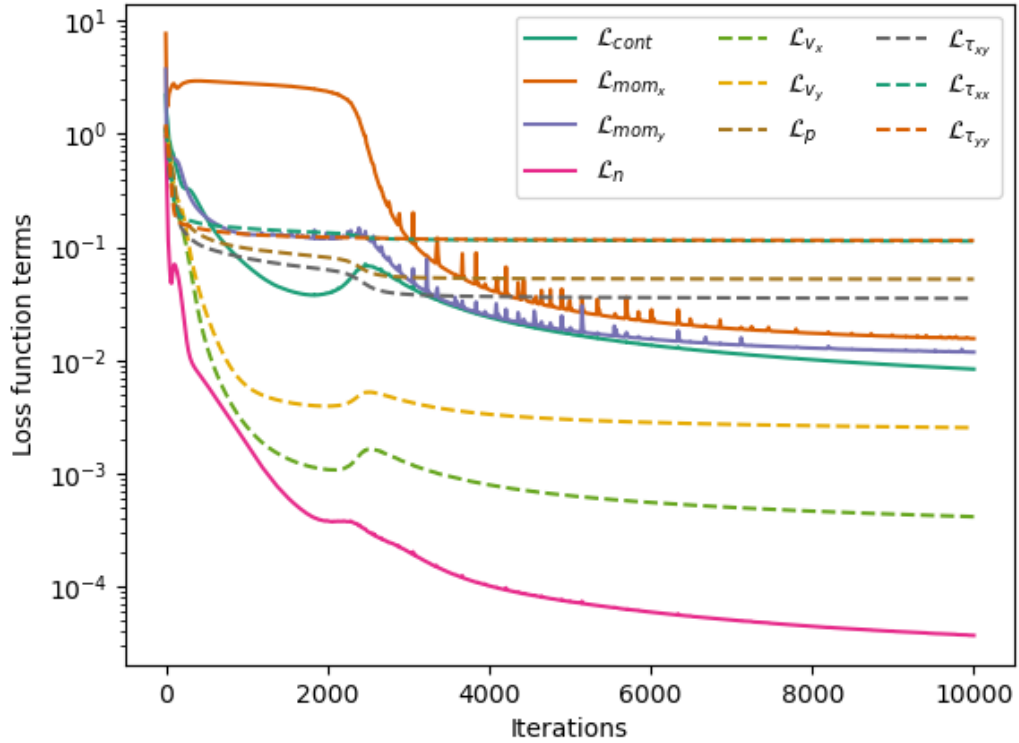

Fig. 9 – Convergence of the neural network during training.

a) 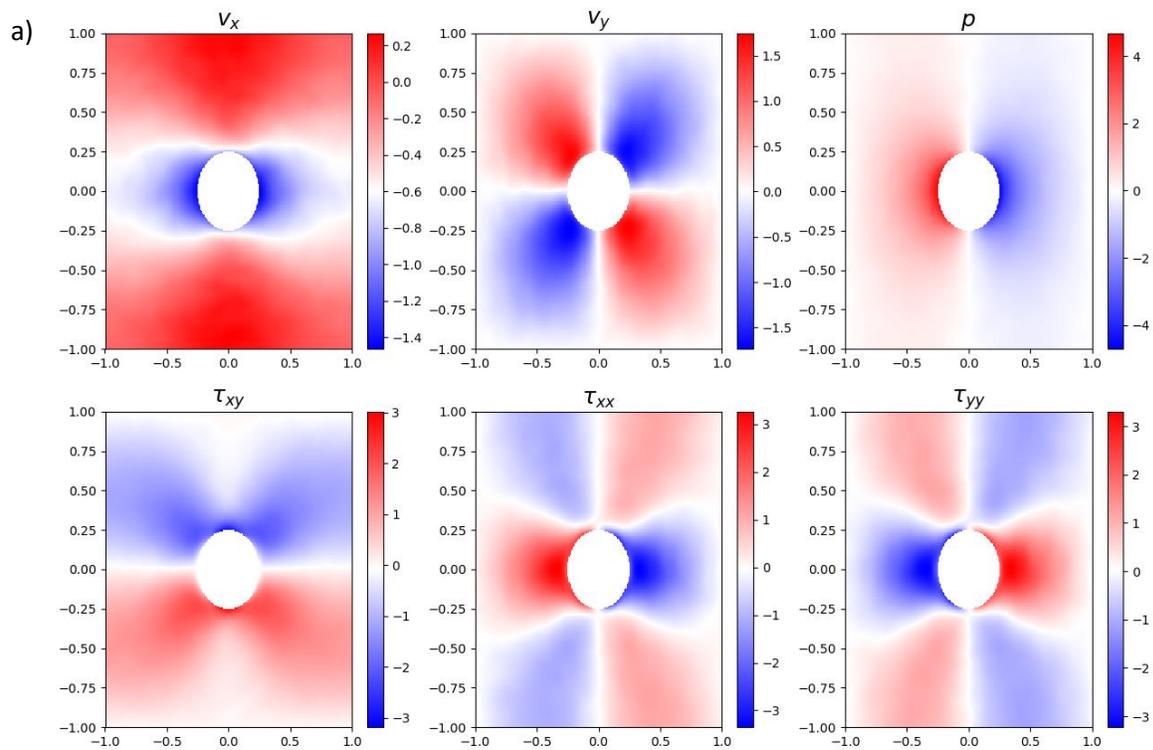

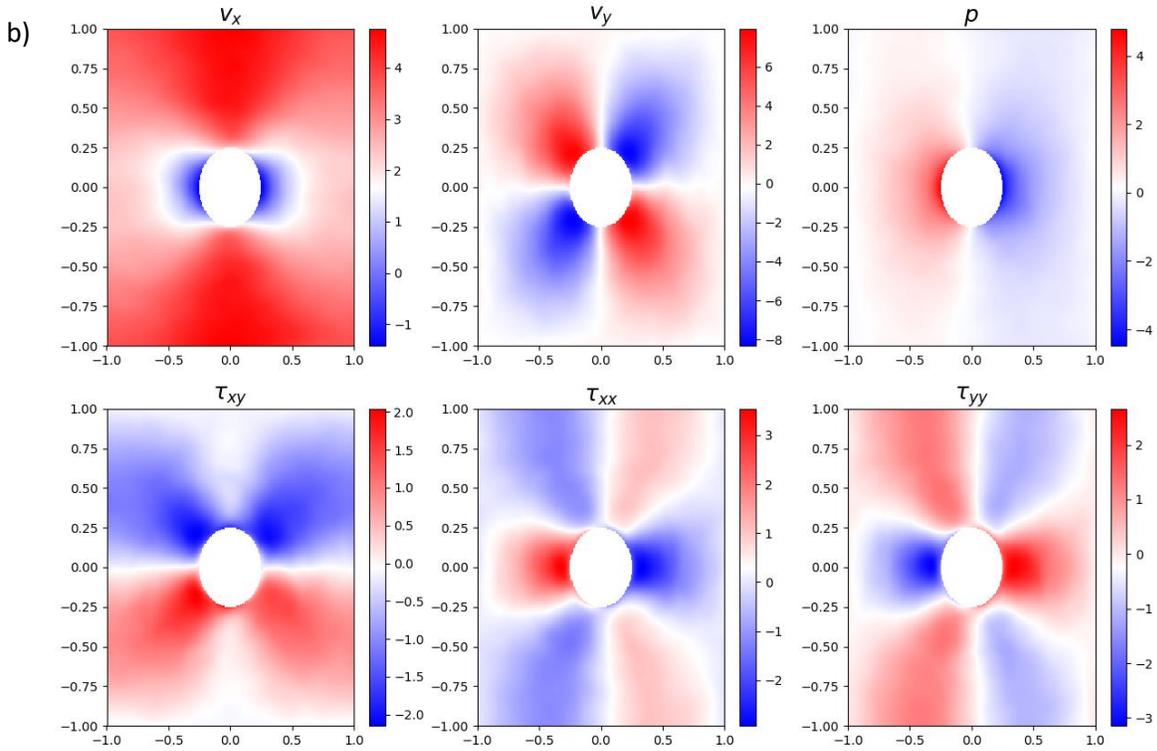

Fig. 10 – PINN predictions of normalized flow variables for a) $Kn = 0.7$, b) $Kn = 5$. Normalization was carried out on the whole dataset, which explains the negative values for $v_x$.

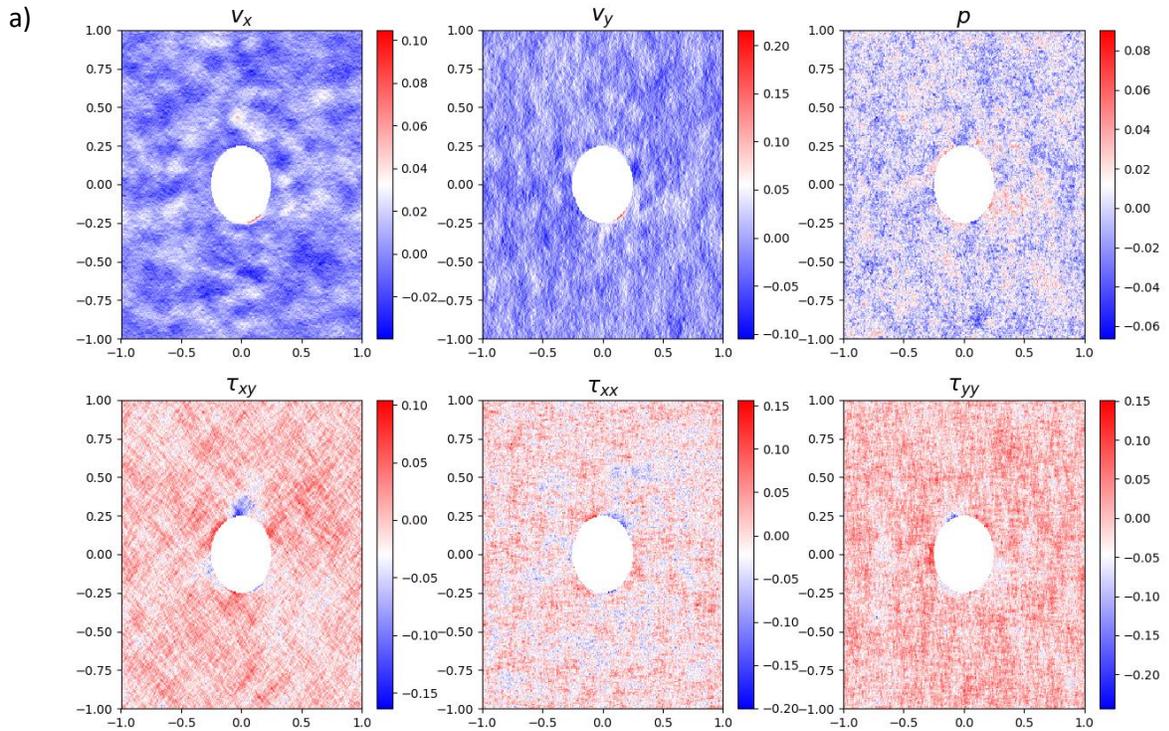

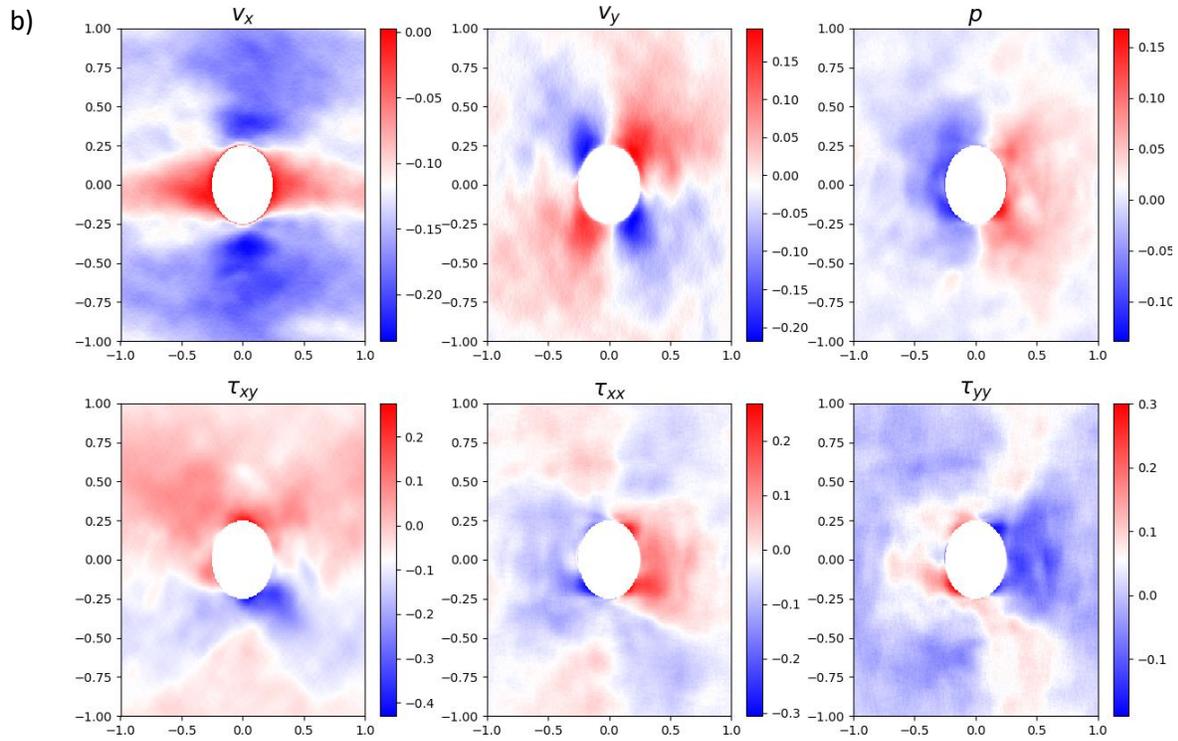

Fig. 11 – Local distance $E$ between dimensionless DSMC fields and the predictions of the PINN for a) $Kn = 0.7$, b) $Kn = 5$. At $Kn = 0.7$, the local distance primarily reflects statistical fluctuations of the original dataset. For $Kn = 5$, however, the distance closely follows the original flow field pattern, indicating a shortcoming in the PINN's extrapolation capability.

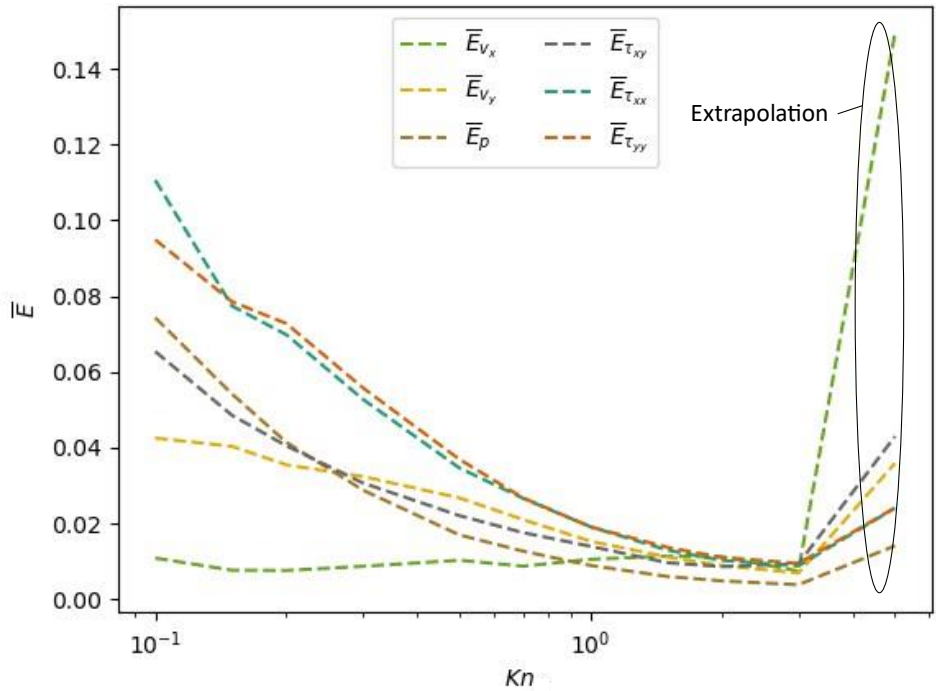

Fig. 12 – Global spatial mean distance as a function of the Knudsen number $Kn$.

### c. Effective viscosity computation

We present here an immediate application of the PINN to qualitatively study rarefied gas flow phenomena. The relationship between the deviatoric stress tensor components and the shear rate of a generalized viscous fluid can be defined as follows:

$$\tau_{xx} = -\mu_{xx}(\mathbf{x}) \cdot \left(2\frac{\partial v_x}{\partial x}\right) \tag{25}$$

$$\tau_{yy} = -\mu_{yy}(\mathbf{x}) \cdot \left(2\frac{\partial v_y}{\partial y}\right) \tag{26}$$

$$\tau_{xy} = -\mu_{xy}(\mathbf{x}) \cdot \left(\frac{\partial v_y}{\partial x} + \frac{\partial v_x}{\partial y}\right) \tag{27}$$

where $\mu_{xx}(\mathbf{x})$, $\mu_{yy}(\mathbf{x})$ and $\mu_{xy}(\mathbf{x})$ represent local viscosities of the fluid.

For a gas in the continuum regime, the stress-shear relationship is linear, yielding a uniform viscosity $\mu_\infty = \mu_{xx}(\mathbf{x}) = \mu_{yy}(\mathbf{x}) = \mu_{xy}(\mathbf{x})$. Deviations among these viscosities would indicate the non-Newtonian behavior of a gas flow. Additionally, viscosity is expected to be positive, meaning the stress should oppose the direction of shear. However, when we evaluate the deviatoric stress components and velocity gradients at sufficiently high Knudsen numbers, the viscosity of the fluid may become negative. To circumvent computational pathologies, we define the following dimensionless regularized effective viscosities $\Psi(\mathbf{x})$ are defined:

$$\Psi_{xx}(\mathbf{x}) = \frac{|\tau_{xx}|}{2\mu\left(\left|\frac{\partial v_x}{\partial x}\right| + \epsilon\right)} \tag{28}$$

$$\Psi_{yy}(\mathbf{x}) = \frac{|\tau_{yy}|}{2\mu\left(\left|\frac{\partial v_y}{\partial y}\right| + \epsilon\right)} \tag{29}$$

$$\Psi_{xy}(\mathbf{x}) = \frac{|\tau_{yy}|}{\mu\left(\left|\frac{\partial v_y}{\partial x}\right| + \left|\frac{\partial u_x}{\partial y}\right| + 2\epsilon\right)} \tag{30}$$

where the small parameter $\epsilon = 0.0001$ is used to avoid division by zero.

Resulting regularized effective viscosities are shown in Fig. 13. In Fig. 13a, $\Psi_{xx}(\mathbf{x}) \approx \Psi_{yy}(\mathbf{x}) \approx 1$. This is expected, as $Kn = 0.1$ is still in the higher end of the slip regime. However, in the band where $-D < x < D$, where shear between cylinders is maximum, $\Psi_{xy}(\mathbf{x})$ is noticeably smaller than zero. This demonstrates the weakness of characterizing a rarefied gas flow through a global Knudsen number $Kn_\infty$, as this phenomenon is clearly linked with the local geometry. By contrast, Fig. 13b shows significant deviations from Newtonian behavior across large subregions of the domain for all regularized effective viscosities, as expected at $Kn = 3$, which is well within the transition regime.

These results challenge of recovering a single effective viscous function [70], that could be applied in a Navier-Stokes solver for non-planar geometries (see [71, 72] for a complete review). However, the PINN model we developed can effectively leverage multidimensional DSMC data mitigating the large noise associated with DSMC's intrinsic statistical fluctuations.

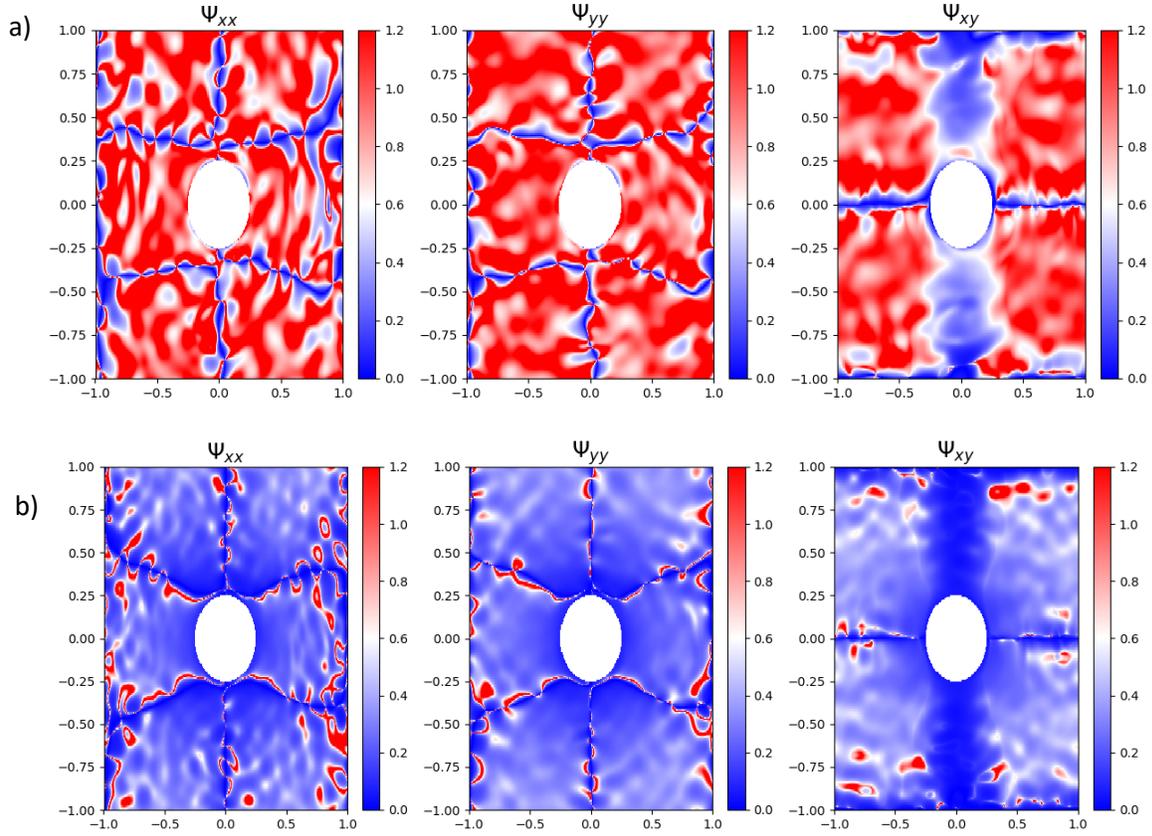

Fig. 13 – Dimensionless regularized effective viscosities for a) $Kn = 0.1$, b) $Kn = 3$.

## 6. Conclusion

In this paper, we developed and tested a new physics-informed neural network designed to predict rarefied gas flows by incorporating the continuity equation and the Cauchy momentum equation as the physics-informed components of the neural network. Using SPARTA, a direct simulation Monte Carlo code, we generated data for flows through a cylinder array on a broad range of Knudsen number ($0.1 < Kn < 5$). We then trained a feed-forward neural network using the DeepXDE platform to approximate the velocity, pressure, and deviatoric stress fields on a data set with $0.1 < Kn < 3$, while withholding $Kn = 0.7$ and $Kn = 5$ to evaluate the PINN's ability to generalize and extrapolate.

Despite statistical fluctuations intrinsic to the DSMC method for weakly compressible laminar flow, our PINN achieved a less than 2% error on the loss function for the continuity and Cauchy momentum equations. Analysis of the difference between SPARTA-generated fields and PINN predictions indicates that the discrepancy mainly arises from the statistical fluctuations present in SPARTA's flow field data. For the excluded test case at $Kn = 0.7$ the model performed well due to interpolating within the training data set. However, at $Kn = 5$, the model exhibited overshoots of up to 20%. Unlike the DSMC results, the PINN predictions were smooth, due to both the activation functions and the intrinsically regularizing properties of the physics-informed contributions to the loss function. While these features enable the PINN to mitigate noise and maintain physical consistency, the overshoots highlight its current limitations in extrapolating beyond the training range.

One key advantage of our PINN framework is its efficiency and simplicity in predicting fewer output variables compared to models relying on higher-order moments or probability distribution functions. In two dimensions, only six outputs are needed: two for velocity, one for pressure, and three for the deviatoric stress tensor. While generating each SPARTA-generated field requires approximately 20 hours on a single node with 4 GPUs, the PINN training process on the entire dataset took less than two hours on one GPU, with model evaluation times under two seconds. These results demonstrate the ability of the PINN to bridge the gap between two simulations with different Knudsen numbers and produce accurate velocity fields with orders-of-magnitude lower computational and memory demands. Extending this approach to three dimensions would require only ten outputs (three for velocity, one for pressure, and six for deviatoric stress tensors) and would likely retain the efficiency and predictive capabilities demonstrated in this study.

While these results are promising, our study has some limitations. We did not conduct a systematic hyperparameter search, and exploring hard boundary conditions on the cylinder surface would be beneficial. Implementing a single Dirichlet boundary condition on the cylinder surface across a broad Knudsen range is complex, unlike the periodic boundary conditions used in our study. Additionally, our PINN results are not immediately generalizable to different geometries; achieving this would require operator learning, which we plan to explore in the near future.

## Acknowledgements

J.-M. T. thanks the FRQNT "Fonds de recherche du Québec – Nature et technologies (FRQNT)'' for financial support (Research Scholarship No. 314328). The authors acknowledge funding from the ERC-PoC2 grant No. 101081171 (DropTrack). The authors are grateful for the CINECA allocation No. 115C.

## Conflict of Interest Statement

The authors declare that they have no known competing financial interests or personal relationships that could have appeared to influence the work reported in this paper.

## Author contributions

**[JMT]**: Conceptualization (equal), Data curation (equal), Formal analysis (equal), Funding acquisition (equal), Investigation (lead), Methodology (equal), Project administration (equal), Resources (equal), Software (equal), Supervision (equal), Validation (equal), Visualization (equal), Writing – original draft, Writing – review & editing (equal); **[ML]**: Formal analysis, Investigation (supporting), Methodology, Resources, Software, Writing – review & editing; **[MD]**: Formal

analysis, Investigation (supporting), Methodology, Software, Writing – review & editing; **[GG]**: Formal analysis, Investigation (supporting), Methodology, Writing – review & editing; **[AM]**: Conceptualization, Formal analysis, Investigation (supporting), Methodology, Project administration, Resources, Supervision, Writing – review & editing; **[SS]**: Conceptualization, Formal analysis, Funding acquisition, Investigation (supporting), Methodology, Project administration, Supervision, Writing – review & editing.

**Data availability statement**

The data that support the findings of this study are available from the corresponding author upon reasonable request.